\documentstyle [12pt] {article} 

\newcommand{\sect}[1]{ \section{#1} } 
 
\newcommand{\req}[1]{(\ref{#1})} 

\newcommand{\ve}{\left( \begin{array}{r}} 
\newcommand{\ev}{\end{array} \right)} 
\newcommand{\ar}{\left( \begin{array}{ll}} 
\newcommand{\ra}{\end{array} \right)} 
\newcommand{\arr}{\left( \begin{array}{rrrr}} 
\newcommand{\arrr}{\left( \begin{array}{rrrrrr}} 

\newcommand{\eqr}{\begin{eqnarray}}
\newcommand{\rqe}{\end{eqnarray}} 
\newcommand{\eq}{\begin{equation}} 
\newcommand{\qe}{\end{equation}}

\newcommand{\half}{\frac{1}{2}}

\newcommand{\eps}{\epsilon}
\newcommand{\paa}{\partial_\alpha}
\newcommand{\pab}{\partial_\beta}

\begin{document}

\begin{flushright}
HUTP-97/A027\\
hep-th/9706087
\end{flushright}

\vspace{2cm}

\begin{center}

{\bf\Large
The Seiberg-Witten Differential From M-Theory
} 

\vspace{1.5cm}

Ansar Fayyazuddin and Micha\l\ Spali\'nski\\
\vspace{0.2cm}

{\em
Lyman Laboratory of Physics\\
Harvard University\\
Cambridge, MA 02138, U.S.A.
}
\end{center}

\vspace{1.4cm}

\begin{abstract} 
The form of the Seiberg-Witten differential is derived from the M-theory 
approach to $N=2$ supersymmetric Yang-Mills theories by directly imposing
the BPS condition for twobranes ending on fivebranes.  The BPS condition
also implies that the pullback of the K\"ahler form onto the space
part of the twobrane world-volume vanishes. 
\end{abstract}

\thispagestyle{empty}

\newpage

\setcounter{page}{1}

\sect{Introduction}

String theory provides a powerful means of analyzing and, at times, solving
supersymmetric field theories.  It also provides a physical interpretation
of the auxiliary geometric structures involved in the solution, which
from the field theory point of view appear in a rather mysterious fashion.
String theory methods were first used to obtain exact results in $N=2$
supersymmetric theories in \cite{kv,kachruetal} using the duality of
heterotic strings on K3$\times T^2$ and type II strings on an appropriate
Calabi-Yau threefold.  It was soon realized that these results only
depend on the local properties of these compactifications, i.e. the ADE
singularities of the K3 fiber, which itself could be replaced by an
appropriate ALE space.  These results were refined further in
\cite{klmvw} where it was realized that one could give a very simple
picture for the existence of BPS saturated states in the field theory as
geodesics around which certain six dimensional self-dual strings could
wrap.  This was taken a step further in \cite{kkv} where field theories
were constructed using only local properties of type IIA on Calabi-Yau
manifolds and performing a local mirror transformation to get the quantum
corrected type IIB dual. Some of these developments are reviewed in
\cite{lerche,klemm}.
 
Another approach which has proven fruitful is to construct quantum field
theories as world-volume theories on branes\cite{bound}.  Theories with
$N=2$ supersymmetry in four dimensions have been constructed in terms of
type IIA fivebranes in \cite{klmvw}. This description was recently
reinterpreted and extended by Witten \cite{wit} in M-theory language in a
way which brings out the connection with world-volume theories of branes
suspended between fivebranes \cite{dia,hw}. The solution of the low energy
dynamics of these $N=2$ supersymmetric quantum field theories
\cite{sw1,sw2} can be obtained by appealing to the $M$-theory
origin\cite{mtheory} of the type IIA string theory.  One of the results of
this point of view is a geometric interpretation of the field theory
beta function. This is the approach which is developed in this note.

The solution of the low-energy dynamics of $N=2$ supersymmetric
Yang-Mills theory involves the so-called Seiberg-Witten differential. As
stressed in \cite{klmvw,war}, in string theory this one-form has a meaning
which goes beyond its function in field theory, where it serves merely as
the object whose periods give the central charges of the $N=2$
supersymmetry algebra (and so in consequence, the relationship between
masses and charges of BPS states). While in field theory the Seiberg-Witten
differential is defined up to exact terms, in string theory a particular
choice of representative is made.  This has led to the
identification of BPS states and geodesics
\cite{klmvw,aj,bs,af,war,rab}. The interpretation of the self-dual string
along the geodesic as the boundary of a twobrane ending on a fivebrane is
the most natural one from the M-theory point of view explored below.

In this note we show how the Seiberg-Witten differential arises in the
M-theory setup proposed in \cite{wit}. The picture which emerges is
actually very similar to that discussed in \cite{klmvw} in a IIA string
theory approach to the N=2 Yang-Mills quantum field theory. In this
approach the relationship between type IIB string theory on an ALE space
near an $A_{n-1}$ singularity and type IIA string theory in the presence of
$n$ symmetric fivebranes was used to give a geometric interpretation of the
Seiberg-Witten differential in the $SU(n)$ case.  One of the chief virtues
of \cite{wit} is to give a simple derivation of results in $N=2$
supersymmetric quantum field theories which were originally obtained in a
more complicated way.  In the same spirit, we give a simple derivation of
the Seiberg-Witten differential by analyzing the BPS condition directly in
M-theory\footnote{For a related discussion see \cite{klmvw,lan} and
\cite{pet}.}. This argument is valid for all classical groups, as it does
not use the explicit form of the Seiberg-Witten curves.

The strategy followed here is to find the BPS states of the $N=2$
supersymmetric gauge theory described as M-theory
twobranes ending on the fivebrane.  The mass
of such a state is given by the membrane tension times the area of the
membrane. The area form can be found by considering the conditions for
preserving supersymmetry, as in \cite{bbs}. The resulting
mass formula can then be
compared to the general expression expected on the basis of the
Seiberg-Witten analysis \cite{sw1,sw2}, which then gives the
Seiberg-Witten differential.

\sect{Embedding $N=2$ supersymmetric Yang-Mills in M-theory}

The four dimensional $N=2$ supersymmetric $SU(n)$ Yang-Mills theory can be
regarded as the world-volume theory of parallel fourbranes in type IIA
theory with finite extent in one direction, $x^6$.  The ends of the
fourbranes, along the $x^6$ axis, lie on fivebranes.  The remaining
coordinates of the world-volume theory of the fourbranes are $x^0, x^1,
x^2, x^3$.  In the remaining dimensions this ``world'' is at a definite
point.

The fivebranes on which the fourbranes end have world-volumes extending
in the $x^0,x^1,x^2,x^3,x^4,x^5$ coordinates, and each is localized at a
point in the remaining coordinates. In the simplest situation there are
just two parallel fivebranes found at two finitely separated points in the
$x^6$ direction\footnote{The picture as described here is accurate only in
a certain semiclassical limit, as discussed in \cite{wit}.}.

Noting that the fourbrane is just an M-theory fivebrane wrapped around
$x^{10}$, one can ask what the M-theory configuration of fivebranes would
have to be to lead to this type IIA configuration of fourbranes and
fivebranes in the weakly coupled limit.  The answer\cite{klmvw,als,wit} is
that in M-theory this configuration can be described simply by a {\em
single} M-theory fivebrane with a world-volume $R^4\times\Sigma$ where the
$R^4$ here is parameterized by coordinates $x^0, x^1, x^2, x^3$ and
$\Sigma$ is a Riemann surface of genus $g=n-1$ embedded in the Euclidean
space $X$ spanned by $x^4,x^5,x^6,x^{10}$.  One can endow $X$ with a
complex structure such that $s= x^6+ix^{10}$ and $v=x^4+ix^5$ are
holomorphic and require that $\Sigma$ is a holomorphic curve in $X$. This
curve is specified by a holomorphic constraint in $v$ and $t \equiv\exp
-s/R$ (this is because the $x^{10}$ coordinate is compactified on a circle
of radius $R$). The form of this constraint can be found explicitly by
realizing that at any value of $v$ there should be two fivebranes
which remain five branes when the radius $R$ of $x^{10}$ is taken to zero.
Furthermore, at any value of $t$ there should be $n$ fourbranes.
These two facts determine $\Sigma$ to be a holomorphic curve $F(v,t)=0$ of
degree $2$ in $t$ and degree $n$ in $v$.  In the coordinates $x^7,x^8,x^9$
the fivebrane is localized at some definite point.

Similar considerations apply to the construction of SO($2n$), SO($2n+1$),
and Sp($n$) theories \cite{als,sasa,lan,bsty}.  The main new ingredient is the
addition of an orientifold fixed plane parallel to the fourbranes.
By working on the covering space this essentially doubles the number
of fourbranes  by introducing the orientifold images of the fourbranes
not localized on the orientifold fixed plane.  The net effect from
the point of view of M-theory is to restrict the form and degree of the
polynomial describing $\Sigma$ to take into account the positions of the
images of the fourbranes.  

In the IIA limit one can include matter hypermultiplets
\cite{wit,lan,sasa,bsty} by introducing additional fourbranes ending on the
``other side'' of the IIA fivebranes.  If the fourbranes whose
world-volume describes the Yang-Mills theory end on the left of a fivebrane
then the matter multiplets end on the right of the fivebrane and vice
versa.  This configuration can similarly be lifted to an M-theory
description by putting appropriate restrictions on the holomorphic curve
$F$ which defines $\Sigma$\cite{wit,lan,sasa,bsty}.

Thus the picture that emerges is that the brane configurations needed to
describe $N=2$ supersymmetric gauge theories can be lifted to M-theory as a
single fivebrane with world-volume $x^0,x^1,x^2,x^3, \Sigma$ with $\Sigma$
a holomorphic curve in $X$.  These cases will be treated in a unified way
below.

\sect{BPS states}

On the fivebrane world-volume there is a two-form whose field strength is
self-dual. The gauge fields in the four dimensional
space $x^0,...,x^3$ are obtained by reducing this self-dual two
form on the Riemann surface $\Sigma$.  This gives rise to $g=n-1$
electric U($1$) fields and $g=n-1$ magnetic U($1$) duals 
\cite{ver,wit}.  BPS states in the fivebrane world-volume theory are 
twobranes ending on the M-theory fivebrane\cite{stromopen,tow}. 
The twobrane boundary has
to lie on the Riemann surface $\Sigma$ if they are to represent
point particles in the $x^0,...,x^3$ space.  The boundary couples to the self-dual
two form on the fivebrane world-volume.  If the twobrane has a 
homologically non-trivial boundary then it will couple to the field theory
gauge fields obtained by reduction of the self-dual 2-form on that
particular homology cycle.

Before analyzing the BPS condition for the twobrane let us discuss 
the supersymmetry preserving condition for a fivebrane wrapped around 
a holomorphic curve.  The condition for supersymmetric $3$-cycles
was given in \cite{bbs} for Calabi-Yau threefold compactifications and has 
been studied in \cite{bsv} in the context
of two branes wrapping two cycles of a K3.

The number of supersymmetries preserved by a p-brane configuration is given
by the number of spinors $\eta$ which satisfy the equation \cite{bbs}
\eq
\label{suco}
\eta = \frac{1}{p!}\eps^{\alpha_1 \dots \alpha_p}
\Gamma_{M_1\dots 
M_p} \partial_{\alpha_1} X^{M_1} \dots \partial_{\alpha_p} X^{M_p}
\eta ,
\qe
where\footnote{The antisymmetrization in \req{muga} includes a factor of
$1/p!$, and the epsilon symbol in \req{suco} is a tensor, not a tensor
density.}  
\eq
\label{muga}
\Gamma_{M_1\dots M_p} = \Gamma_{[M_1}\dots\Gamma_{M_p]}
\qe
and  $X^{M}$ is the embedding of the p-brane in $R^{9,1}\times S^1$.  
Consider first the M-theory fivebrane configuration. For a fivebrane with
world-volume 
filling $x^0,....,x^4,\Sigma$ the supersymmetry condition (in the static
gauge) reduces to:  
\eq
\eta = \frac{1}{2}\eps^{\alpha \beta}
\Gamma_{0}\dots \Gamma_{3} \Gamma_{ij} 
\partial_{\alpha} X^i \partial_{\beta} X^j \eta ,
\qe
where $i,j$ label the coordinates $(X^4,X^5,X^6,X^{10})$. 
At this point it is useful to pass to complex coordinates, defined by
\eqr
S&=&X^4+iX^5 ,\\
V&=&X^6+iX^{10}
\rqe
which we will denote by $(X^m,X^{\bar{m}})$. 
\eq
\eta = \frac{1}{2}\eps^{\alpha \beta}
\Gamma_{0}\dots \Gamma_{3} (\Gamma_{mn} 
\partial_{\alpha} X^m \partial_{\beta} X^n + \Gamma_{m{\bar n}} 
\partial_{\alpha} X^m \partial_{\beta} X^{\bar n} + 
\Gamma_{{\bar m}{\bar n}} \partial_{\alpha} X^{\bar m} \partial_{\beta} 
X^{\bar n}) \eta.
\qe
As the fivebrane is wrapped around the holomorphic curve $\Sigma$, only
the term with one holomorphic and one anti-holomorphic index is non-vanishing:
\eq
\eta = \frac{1}{2}\eps^{\alpha\beta}
\Gamma_{0}\dots \Gamma_{3} \Gamma_{m\bar{n}} 
\partial_{\alpha} X^m \partial_{\beta} X^{\bar{n}}
\eta.
\qe
From this it follows that
\eq
\Gamma_{0}\dots \Gamma_{3} \Gamma_{m\bar{n}} \eta = J_{m\bar{n}} \eta ,
\qe
where $J$ 
is the K\"ahler form on $X$ ($J_{m{\bar n}}=ig_{m{\bar n}}$). This gives 
\eqr
\label{spin51}
i\Gamma_{0}\dots \Gamma_{3} \Gamma_{V\bar{V}} \eta &=& \eta ,\nonumber \\ 
i\Gamma_{0}\dots \Gamma_{3} \Gamma_{S\bar{S}} \eta &=& \eta ,\nonumber \\
\Gamma_{V\bar{S}} \eta &=& 0 ,\nonumber \\
\Gamma_{S\bar{V}} \eta &=& 0 .
\rqe
The $11$-dimensional spinor $\eta$ is a $32$-component complex spinor,
which satisfies the Majorana condition \footnote{See the appendix for the
definition of $B$.} 
\eq
\eta = B\eta^\star .
\qe
It can thus be written as 
\eq
\eta = \chi + B\chi^\star .
\qe
Using 
\eq
B\Gamma_m =-\Gamma_{\bar m} B,
\qe
which follows from the expressions given in the appendix, one finds
\eqr
\label{annih}
\Gamma_{\bar{V}} \chi &=& 0 \nonumber\\
\Gamma_{\bar{S}} \chi &=& 0 .
\rqe
The surviving $\eta$ has $32/4 = 8$ complex components. From \req{spin51} it then
follows that
\eq
\label{chir}
i\Gamma_{0}\dots \Gamma_{3} \chi = - \chi ,
\qe
which cuts the number of solutions by another factor of 2, so one concludes
that there are $4$ complex solutions, which confirms that 
the fivebrane configuration under consideration indeed gives $N=2$
supersymmetry in the four dimensional sense. 

Introducing the twobrane which ends on the fivebrane requires 
the additional constraint\cite{stromopen}:
\eq
\eta = \frac{1}{2}\eps^{\alpha\beta}
\Gamma_{0} \Gamma_{ij} 
\paa X^{i} \pab X^{j}
\eta ,
\qe
where $X$ now denotes the embedding of the twobrane. Taking into account
equations \req{spin51}, one finds
\eqr
\label{spin52}
\eps_{\alpha\beta} \eta &=& 
\Gamma_{0} [
(\paa S \pab V - \paa V\pab S) \Gamma_{SV} + 
(\paa \bar{S} \pab \bar{V} - \paa \bar{V}\pab \bar{S})
\Gamma_{\bar{S}\bar{V}} \nonumber \\ 
&-& i (\paa S \pab \bar{S} + \paa V \pab \bar{V} - \paa \bar{S}\pab \bar{S} -
\paa \bar{V}\pab V) \Gamma_0\dots  \Gamma_3 ] \eta .
\rqe
To analyze this equation it is convenient to define the following
projection operators:
\eqr
P_+ &=& (1+i\Gamma_0 \dots \Gamma_3)/2 ,\nonumber \\
P_- &=& (1-i\Gamma_0 \dots \Gamma_3)/2
\rqe 
and 
\eqr
Q_+ &=& \Gamma_S\Gamma_{\bar{S}}\Gamma_V\Gamma_{\bar{V}} ,\nonumber
\\
Q_- &=& (1-\Gamma_S\Gamma_{\bar{S}}\Gamma_V\Gamma_{\bar{V}}) .
\rqe
These projection operators satisfy a set of simple relations which follow
directly from their definition:\footnote{The relevant properties of the $11$
dimensional Dirac algebra are collected in the appendix.}
\eq
P_\pm \Gamma = \Gamma P_\mp \ , \qquad \Gamma\in \{\Gamma_V, \Gamma_S,
\Gamma_{\bar{S}}, \Gamma_{\bar{V}}\}
\qe
and
\eqr
P_+ B &=& B P_- , \qquad P_+ \Gamma_0 = \Gamma_0 P_- ,\nonumber \\
Q_+ B &=& B Q_- , \qquad Q_\pm \Gamma_0 = \Gamma_0 Q_\pm .
\rqe
From \req{chir} it follows that 
\eqr
P_+\chi &=& 0, \qquad P_+ B\chi^\star = B\chi^\star ,\nonumber \\
P_-\chi &=& \chi, \qquad P_-B\chi^\star = 0
\rqe
and from \req{annih} one finds:
\eqr
Q_+\chi &=& 0, \qquad Q_+ B\chi^\star = B\chi^\star ,\nonumber \\
Q_-\chi &=& \chi, \qquad Q_- B\chi^\star = 0.
\rqe
Acting on \req{spin52} with $Q_+$ gives\footnote{Projection with $Q_-$ leads to
the conjugate equations.}
\eqr
\label{spin22}
\eps_{\alpha\beta} B\chi^\star &=& (\paa S\pab V - \paa V\pab S)
\Gamma_0\Gamma_S\Gamma_V\chi 
\nonumber \\ 
&-& i(\paa S\pab \bar{S} + \paa V\pab \bar{V}- \paa \bar{S}\pab S- \paa
\bar{V}\pab V) \Gamma_0 B\chi^\star .
\rqe
One can now project the $P_+$ piece from \req{spin22}, which implies
\eq
\label{mem}
B\chi^\star = \Gamma_0\Gamma_S\Gamma_V\chi 
\qe
and that the volume form on the spatial part of the two brane world-volume
is the pullback of the holomorphic 2-form:
\eq
\omega = ds\wedge dv .
\qe
Projecting with $P_-$
leads to the conclusion that the pullback of the K\"ahler form to the
twobrane must vanish (a similar result was found in a different context in
\cite{bbs}):
\eq
\paa S\pab \bar{S} + \paa V\pab \bar{V}- \paa \bar{S}\pab S- \paa
\bar{V}\pab V =0 .
\qe

Equation \req{mem} cuts the number of supersymmetries by half,
expressing the fact that the membrane state considered here breaks half of
the supersymmetry (leaving 4 real supersymmetries). Thus this membrane is a
BPS state in the world-volume theory of the fivebrane.

\sect{The Seiberg-Witten Differential}

As explained in the introduction, the Seiberg-Witten differential can be
inferred from the formula for the mass of a BPS state. 

The mass of a BPS saturated twobrane is given by the brane tension times
the area:
\eq
m = T_2\times (\mbox{area})= T_2 \int_{M_2} \mid\omega\mid .
\qe
The last equality follows from the analysis in the previous section
where it was found that the area form on the twobrane world-volume is the
pullback of the holomorphic two form $\omega$.
Using the fact that the twobrane has a boundary on $\Sigma$ one can
use Stokes' theorem to write the mass as an integral on the
boundary, which in terms of $t\equiv e^{-s}$ reads
\eq
m^2 = T_2 \mid\int_{\partial M_2} v(t) \frac{dt}{t}\mid^2 .
\qe
Here $v(t)$ is $v$ written as a function of $t$ on $\Sigma$, since
the boundary $\partial M_2$ lies on $\Sigma$.
Thus the mass of the twobrane reduces to an integral of a meromorphic
1-form on the Riemann surface $\Sigma$.  From this equation it follows
that the Seiberg-Witten differential is:
\eq
\lambda_{SW} = v(t)\frac{dt}{t}.
\qe

The form of the Seiberg-Witten differential obtained above is the one which
naturally appears in the integrable systems approach $N=2$ supersymmetric
Yang-Mills \cite{mw}.

\sect{Conclusions} 

The analysis of the BPS conditions for a twobrane ending
on a fivebrane in the M-theory approach to $N=2$ supersymmetric gauge
theories leads to two conditions on the twobrane.  The first one is that
the induced K\"ahler form on the twobrane vanish, and the second is that
the pullback of the holomorphic 2-form $\omega$ be equal to the area of the
twobrane.  The latter condition leads directly to the Seiberg-Witten
differential. It should also be possible to derive the Seiberg-Witten
differential by directly studying the conditions for world-volume
supersymmetry in the fivebrane field theory\footnote{This was pointed out
to us by C. Vafa.}.

\sect{Appendix}

This appendix presents some facts concerning the Dirac algebra in $11$
dimensions, which are needed to perform the calculations presented in this
paper.

The eleven dimensional metric is taken to have signature $(-,+\dots +)$.

A representation for the Dirac matrices which is convenient here is given
in terms of the following tensor product representation
\eqr
\Gamma_\mu &=& \gamma_\mu \otimes 1 \otimes 1 \qquad \mu=0\dots 3\nonumber \\
\Gamma_{i+3} &=& \gamma_5 \otimes \tilde{\gamma}_i \otimes 1 \qquad
i=1\dots 3 \nonumber \\
\Gamma_{a+6} &=& \gamma_5 \otimes \tilde{\gamma}_5 \otimes \sigma_a \qquad
a=1\dots 3 \nonumber \\
\Gamma_{10} &=& \gamma_5 \otimes \tilde{\gamma}_4 \otimes 1  ,
\rqe
where the first and second factors are four-dimensional and the last one is
two-dimensional. The individual factors are 
\eq
\gamma_\mu = i \ar 0 & \sigma_\mu \\ \bar{\sigma}_\mu & 0 \ra
\qe
and $\sigma^\mu = (1, \sigma^1, \sigma^2, \sigma^3)$ and 
$\bar{\sigma}^\mu = (1, -\sigma^1, -\sigma^2, -\sigma^3)$, the $\sigma^i$
being Pauli 
matrices. Thus 
$\gamma^0, \gamma^1, \gamma^3$ are imaginary, while $\gamma^2$ is
real. Furthermore 
\eq
\gamma_5 = i\gamma_0\gamma_1\gamma_2\gamma_3
\qe 
and
\eqr
\tilde{\gamma}_i &=& \gamma_i \qquad i=1\dots 3; \nonumber \\
\tilde{\gamma}_4 &=& i \gamma_0\nonumber \\
\tilde{\gamma}_5 &=& \gamma_1\gamma_2\gamma_3\gamma_4 .
\rqe
These choices imply that $\Gamma_M$ are real for $M=2,5,7,8,10$, purely
imaginary for $M=0,1,3,4,6,9$, symmetric for $M=0,2,5,7,8,10$ and
antisymmetric for $M=1,3,4,6,9$. 

The charge conjugation matrix is given by
\eq
C=B\Gamma_0 ,
\qe
where the unitary matrix $B$ satisfies
\eq
\Gamma_M^\star = B\Gamma B^{-1}
\qe
and in $11$ dimensions one can choose
\eqr
B &=& B^\star\\
B &=& B^\dagger .
\rqe
In the representation chosen above for the Dirac matrices, we have
\eq
B=\Gamma_2 \Gamma_5 \Gamma_7 \Gamma_8 \Gamma_{10} .
\qe
The text refers to the matrices
\eqr
\Gamma_V &=& \half (\Gamma_4 + i \Gamma_5) \qquad  \Gamma_S =  \half (\Gamma_6 + i \Gamma_{10}) \nonumber \\
\Gamma_{\bar{V}} &=&  \half (\Gamma_4 - i \Gamma_5)  \qquad
\Gamma_{\bar{S}} =  \half (\Gamma_6 - i\Gamma_{10})
\rqe
In the representation chosen here these are purely imaginary. 

\vspace{0.5cm}

\begin{center}
  {\bf Acknowledgments}
\end{center}

We would like to thank C. Vafa for helpful comments.  A.F. would also like
to thank N. Sasakura and D.J. Smith for valuable discussions.  The work of
M.S. was supported by a Fulbright Fellowship.

\newpage               

\newcommand{\bi}[1]{\bibitem{#1}}

\newcommand{\plb}[3]{{\em {Phys. Lett.}} {\bf B {#1}} (19{#2}) {#3}} 
\newcommand{\npb}[3]{{\em {Nucl. Phys.}} {\bf B {#1}} (19{#2}) {#3}} 
\newcommand{\prl}[3]{{\em Phys. Rev. Lett.} {\bf {#1}} (19{#2}) {#3}} 
\newcommand{\cmp}[3]{{\em {Comm. Math. Phys.}} {\bf {#1}} (19{#2}) {#3}} 
\newcommand{\mpla}[3]{{\em Mod. Phys. Lett.} {\bf A {#1}} (19{#2}) {#3}.}
\newcommand{\ijmpa}[3]{{\em Int. J. Mod. Phys.} {\bf A {#1}} (19{#2}) {#3}}


\begin{thebibliography}{99} 

\bi{kv}S. Kachru and C. Vafa,  ``Exact Results for N=2 Compactifications of
Heterotic Strings'', \npb{450}{95}{69}, (hep-th/9505105).

\bi{kachruetal} S. Kachru, A. Klemm, W. Lerche, P. Mayr, C. Vafa,
``Nonperturbative Results on the Point Particle Limit of N=2 Heterotic
String Compactifications'', \npb{459}{96}{537}, (hep-th/9508155).

\bi{klmvw} A. Klemm, W. Lerche, P. Mayr, C. Vafa and N. Warner, ``Self-Dual
Strings and N=2 Supersymmetric Field Theory'' \npb{477}{96}{746}
(hep-th/9604034).

\bi{kkv}  S. Katz, A. Klemm, C. Vafa, ``Geometric Engineering of Quantum
Field Theories'', hep-th/9609239.

\bibitem{lerche} W.\ Lerche, ``Introduction to Seiberg-Witten and its
Stringy Origin'', hep-th/9611190.   


\bi{klemm} A. Klemm, ``On the Geometry behind N=2 Supersymmetric
Effective Actions in Four Dimensions,'' hep-th/9705131.

\bi{bound} E. Witten, ``Bound States Of Strings And $p$-Branes'', \npb
{460}{96}{335} (hep-th/9510135).

\bi{wit} E. Witten, ``Solutions Of Four-Dimensional Field
Theories Via M-Theory,'', hep-th/9703166.

\bi{dia} D. Diaconescu, ``D-branes, Monopoles and Nahm Equations'',
hep-th/9608163.


\bi{hw} A. Hanany, E. Witten, `` Type IIB Superstrings, BPS Monopoles, And
Three-Dimensional Gauge Dynamics'', hep-th/9611230.


\bi{sw1} N. Seiberg and E. Witten, ``Monopole Condensation, and Confinement
in $N=2$ Supersymmetric Yang-Mills Theory,'' \npb {426}{94}{19} (hep-th
9407087).

\bi{sw2} N. Seiberg and E. Witten, ``Monopoles, Duality and Chiral Symmetry
Breaking in $N=2$  Supersymmetric QCD,'' \npb {431}{94}{484} (hep-th/9408099). 

\bi{mtheory} For a review see P.K. Townsend, ``Four Lectures on M-theory'',
hep-th/9612121. 


\bi{war} J. Schulze, N. P. Warner, ``BPS Geodesics in N=2
Supersymmetric Yang-Mills Theory'', hep-th/9702012.


\bi{aj} A. Johansen,  ``A Comment on BPS States in F-theory in 8
Dimensions'', \plb{395}{97}{36} (hep-th/9608186).

\bi{bs} A.\ Brandhuber and S.\ Stieberger, ``Self-Dual Strings and
Stability of BPS States in N=2 SU(2) Gauge Theories'', \npb{488}{97}{199}
(hep-th/96011016).  

\bi{af}A. Fayyazuddin, ``Results in susy field theory from 3-brane probe
in F-theory'',  hep-th/9701185, to appear in Nucl. Phys. B.


\bi{rab} J. M. Rabin, ``Geodesics and BPS States in N=2 Supersymmetric
QCD'', (hep-th/9703145). 

\bi{lan} K. Landsteiner, E. Lopez, D. A. Lowe, ``$N=2$ supersymmetric
Gauge Theories, Branes and Orientifolds'', hep-th/9705199.


\bi{pet} S. Katz, P. Mayr, C. Vafa, ``Mirror symmetry and Exact Solution of
4D N=2 Gauge Theories I'', hep-th/9706110.

\bi{bbs} K. Becker, M. Becker and A. Strominger, ``Fivebranes, Membranes
and Non-perturbative String Theory''\npb{456}{95}{130} (hep-th/9509175). 


\bi{als} Nick Evans, Clifford V. Johnson, Alfred D. Shapere,
     ``Orientifolds, Branes, and Duality of 4D Gauge Theories'',
     hep-th/9703210. 

\bi{sasa} N. Sasakura and D.J. Smith, unpublished.

\bi{bsty} A. Brandhuber, J. Sonnenschein, S. Theisen, S. Yankielowicz
``M-theory and Seiberg-Witten Curves: Orthogonal and Symplectic Groups'',
hep-th/9705232.

\bi{ver} E. Verlinde, ``Global Aspects of Electric-Magnetic Duality'',
\npb{455}{95}{211}, (hep-th/9506011).

\bi{stromopen} A. Strominger, ``Open P-Branes'', \plb {B383}{96}{44},
(hep-th/9512059).

\bi{tow} P. Townsend, ``D-Branes from M-Branes'' \plb
{B373}{96}{68}, (hep-th/9512062).

\bi{bsv} M. Bershadsky, V. Sadov and C. Vafa, ``D-Branes and Topological
Field Theories'', \npb {463}{96}{420}, (hep-h/9511222

\bi{mw}A. Gorsky, I. Krichever, A. Marshakov, A. Mironov, A. Morozov,
``Integrability and Seiberg-Witten Exact Solution'',\plb{355}{95}{466},
(hep-th/9505035);\\
A.\ Marshakov, A.\ Mironov and A. Morozov, hep-th/96071109;\\
E. Martinec and N. Warner, ``Integrable systems and supersymmetric
gauge theory,'' \npb{459}{96}{97} (hep-th 9509161); \\
R. Donagi, E. Witten ``Supersymmetric Yang-Mills Systems And Integrable
Systems'', \npb{460}{96}{299} (hep-th/9510101).


\end{thebibliography}
\end{document}